\documentclass[pra,aps,twocolumn,nopacs,superscriptaddress,nofootinbib]{revtex4}

\usepackage{graphicx}
\usepackage{dcolumn}
\usepackage{bm}
\usepackage{bbm}
\usepackage{amsmath}
\usepackage{epsfig}
\usepackage{indentfirst}
\usepackage{psfrag}
\usepackage{subfigure}
\usepackage{amssymb}
\usepackage{color}
\usepackage[colorlinks,linkcolor=blue,citecolor=blue,urlcolor=blue,hyperindex,driverfallback=dvipdfm]{hyperref}
\usepackage[T1]{fontenc}
\usepackage{comment}
\usepackage{upgreek}

\newcommand{\abs}[1]{\mathopen{}\left|#1\right|\mathclose{}}

\def\ii{{\rm i}}  \def\ee{{\rm e}}
  
  \def\Rb{{\bf R}}    \def\vb{{\bf v}}

  \def\kpar{k_\parallel}  
\def\Qb{{\bf Q}}    
\def\me{m_{\rm e}}  

\begin{document}
\title{Electron Beam Aberration Correction Using Optical Near Fields}
\author{Andrea~Kone\v{c}n\'{a}}
\affiliation{ICFO-Institut de Ciencies Fotoniques, The Barcelona Institute of Science and Technology, 08860 Castelldefels (Barcelona), Spain}
\author{F.~Javier~Garc\'{\i}a~de~Abajo}
\email[Corresponding author: ]{\\ javier.garciadeabajo@nanophotonics.es}
\affiliation{ICFO-Institut de Ciencies Fotoniques, The Barcelona Institute of Science and Technology, 08860 Castelldefels (Barcelona), Spain}
\affiliation{ICREA-Instituci\'o Catalana de Recerca i Estudis Avan\c{c}ats, Passeig Llu\'{\i}s Companys 23, 08010 Barcelona, Spain}

\begin{abstract}
The interaction between free electrons and optical near fields is attracting increasing attention as a way to manipulate the electron wave function in space, time, and energy. Relying on currently attainable experimental capabilities, we design optical near-field plates to imprint a lateral phase on the electron wave function that can largely correct spherical aberration without the involvement of electric or magnetic lenses in the electron optics, and further generate on-demand lateral focal spot profiles. Our work introduces a disruptive and powerful approach toward aberration correction based on light-electron interactions that could lead to compact and versatile time-resolved free-electron microscopy and spectroscopy.
\end{abstract}
\date{\today}
\maketitle



\section{Introduction}

The development and widespread use of spatial light modulators have revolutionized optics by enabling an increasing degree of control over light beam propagation. Likewise, the extension of this concept to electron optics could provide the means for controlling the electron wave function and its interactions with atomic-scale samples. Electron microscopes already reach precise spatial and temporal control over the amplitude and phase of the wave function of beam electrons employed as sample probes. Over the last decades, costly and sophisticated arrangements of magnetostatic and electrostatic lenses have been engineered to eliminate electron optics aberrations \cite{HS19}, making it possible to focus electron beams with sub-{\AA}ngstrom precision in state-of-the-art scanning transmission electron microscopes. These capabilities are crucial for atomic-scale imaging and spectroscopy \cite{BDK02,MKM08}.

Parallel efforts have led to the development of amplitude and phase reconstruction techniques such as ptychography \cite{JCH18} or electron tomography and holography \cite{MD09}, which have proved useful in both imaging low-contrast samples \cite{FSE97,YHC18} and acquiring additional information on the sample, such as electric or magnetic field distributions \cite{MAC10,NPL13,SKK17}. An alternative approach have consisted in preparing electron beams with on-demand focal spot phase and intensity distributions designed to introduce phase contrast and selectively interact with targeted types of excitations such as plasmons of specific multipolar symmetry \cite{GBL17} or chiral modes and materials magnetic properties \cite{LBT17,RB13_2}. Such phase-shaped electron beams can be obtained through diffraction by a static phase plate \cite{VTS10,MAA11,SLL14,GBL17} or ingenious use of lens aberrations \cite{SSV12,CBK13}. A recent work also demonstrates programmable electron phase plates based on arrays of electrically biased transmission elements \cite{VBM18}.


The interaction of free electrons with optical near fields in illuminated nanostructures opens exciting possibilities as a further mechanism to control the electron wave function. This phenomenon has been exploited to develop the so-called photon-induced near-field electron microscopy (PINEM) \cite{BFZ09}, which has been the subject of intense experimental \cite{BFZ09,KGK14,PLQ15,FES15,paper282,EFS16,KSE16,RB16,VFZ16,MB17,KML17,FBR17,PRY17,paper306,paper311,MB18,paper332,WDS19,DNS19,KLS19} and theoretical \cite{paper151,PLZ10,PZ12,paper272,paper312,paper339,K19,RML19} efforts. By synchronizing the arrival of ultrashort electron and laser pulses near the sample, the former can undergo stimulated absorption or emission of up to 100s of photons \cite{DNS19,KLS19}. This technique has been predicted to imprint optical phase on the lateral electron wave function \cite{paper312}, which has been demonstrated to generate vortex electron beams via photon-to-electron angular momentum transfer \cite{paper332}. The synergetic combination of spatial light modulators and ultrafast electron microscopy constitutes a powerful platform for the control of free-electron wave functions, including the possibility of compensating beam aberrations and shaping the focal spot, as an alternative to traditional electron optics components. 

\begin{figure}
\centering \includegraphics[width=0.375\textwidth]{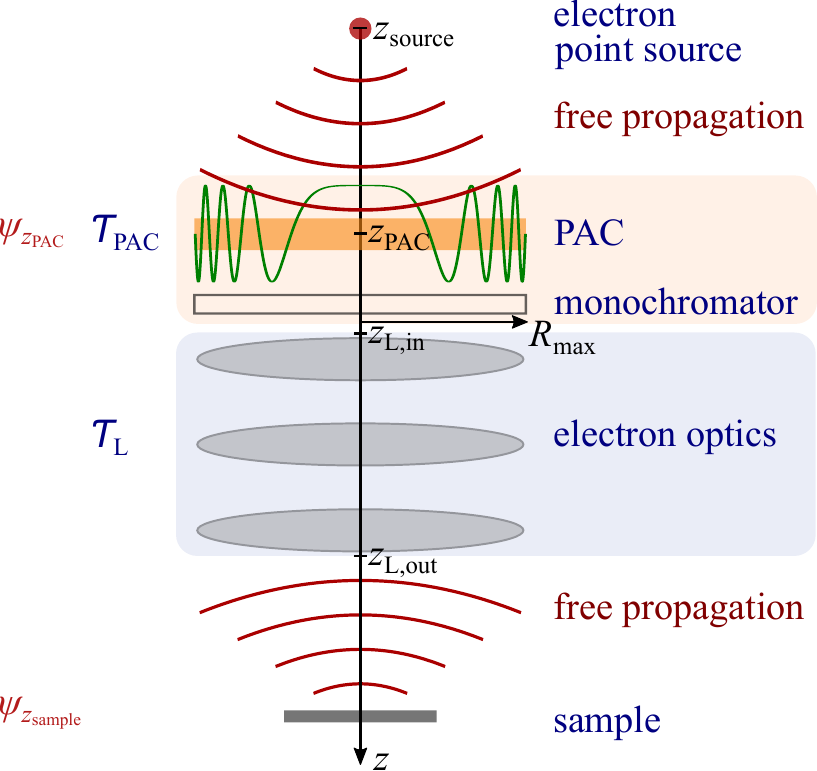}
\caption{Proposed experimental arrangement incorporating a photonic aberration corrector (PAC) to mitigate electron spherical aberration through an electron optical phase plate. The PAC module (light orange frame) is placed just before the electron optics focusing module (electromagnetic lenses inside the light blue frame) at a distance $z_{\rm L,in}-z_{\rm source}$ from the electron source. The transmitted beam is restricted by a circular aperture of radius $R_\mathrm{max}$.}
\label{Fig1}
\end{figure}

In this work, we theoretically demonstrate the correction of spherical aberration in an electron beam upon transmission through an illuminated thin film, where light-electron phase transfer compensates for the undesired deviation of the transversal electron wave function from a spherical wavefront, thereby resulting in nearly unaberrated focusing down to sub-{\AA}ngstrom focal spots. The proposed implementation of this type of photonic aberration corrector (PAC) in an electron microscope is schematically depicted in Fig.~\ref{Fig1}, where the new element is placed after the electron source in order to imprint the required phase on the electron wave function to compensate for the aberration produced by subsequent electron-optics focusing elements. The PAC consists of an optically-opaque electron-transparent thin film (e.g., a metal film deposited on a Si$_3$N$_4$ membrane, as already used in PINEM studies \cite{paper282,paper311}) on which a lateral optical pattern is projected with diffraction-limited spatial resolution (see Appendix\ \ref{AS3}). Electron interaction with semi-infinite light fields \cite{paper311} in this film then produces energy sidebands in the transmitted electrons that can be optimized to accommodate $\sim1/3$ of the electrons in the first sideband (i.e., electrons that have gained one photon energy). A monochromator inserted right after the PAC removes the rest of the energy sidebands before entering an electron optics module for focusing at the sample. Aberration correction is thus performed through the PAC phase plate, without involvement of complex electron optics. This type of design inherits the flexility of light patterning through spatial light modulators, here demonstrated for aberration correction, but also enabling arbitrary shaping of the electron focal spot.


\section{Electron beam propagation through the electron microscope}

We represent fast electrons by their space- and time-dependent wave function $\psi_z(\Rb)\ee^{-\ii\mathcal{E}_0t}$, where we consider monochromatic electrons of energy $\mathcal{E}_0$ that depend on transversal coordinates $\Rb=(x,y)$ at each propagation plane determined by $z$. Free propagation from $z'$ to $z$ is described by the expression \cite{paper272,AOP01}
\begin{align}
\psi_z(\Rb)=\!\int\!\!\!\int\frac{d^2\Qb d^2\Rb'}{(2\pi)^2}\,\ee^{\ii[\Qb\cdot(\Rb-\Rb')+q_z(z-z')]}
\psi_{z'}(\Rb'),\nonumber
\end{align}
where the outer integral extends over transversal wave vectors $\Qb$, $q_z=\sqrt{q_0^2-Q^2}$ is the longitudinal wave vector, $\hbar{\bf q}_0=\me\vb\gamma$ and $\vb$ are the average electron momentum and velocity vectors, respectively, $\gamma=1/\sqrt{1-v^2/c^2}$ is the Lorentz factor, $\me$ is the electron rest mass, and $c$ is the speed of light in vacuum. In what follows, we focus on electron beams of well-defined chirality, characterized by an azimuthal orbital quantum number $m$, such that the wave function takes the form $\psi_z(\Rb)=\psi_z(R)\ee^{\ii m\varphi}$. Additionally, electrons in microscopes are collimated and therefore safely described in the paraxial approximation $q_z\approx q_0-Q^2/2q_0$. These considerations allow us to carry out some of the above integrals to find (see Appendix\ \ref{AS1})
\begin{align}
&\psi_z(R)=(\mathcal{F}^m_{z-z'}\cdot\psi_{z'})\big{|}_R \label{propagation}\\
&\equiv(-\ii)^{m+1}\xi_{z-z'}\,\ee^{\ii q_0 (z-z')} \nonumber\\
&\quad\times\int_0^\infty \!\!R'dR'\, J_m\left(\xi_{z-z'}RR'\right)\, \ee^{\ii \xi_{z-z'}(R^2+R'^2)/2}\psi_{z'}(R'),
\nonumber
\end{align}
where $\xi_{z-z'}=q_0/(z-z')$, $J_m$ is a Bessel function, and we implicitly define the free-propagation operator $\mathcal{F}^m$ using matrix notation with a dot standing for integration over the radial coordinate $R$.

Transmission through the microscope sketched in Fig.~\ref{Fig1} results in an electron wave function at the sample given by \cite{HLL1995,E05}
\begin{align}
&\psi_{z_{\rm sample}} \label{micros}\\
&=    \mathcal{F}^m_{z_{\rm sample}-z_{\rm L,out}}
\cdot \mathcal{T}_{\rm L}
\cdot \mathcal{F}^m_{z_{\rm L,in}-z_{\rm PAC}} 
\cdot \mathcal{T}_{\rm PAC}
\cdot \psi_{z_{\rm PAC}},
\nonumber
\end{align}
where $\psi_{z_{\rm PAC}}$ represents the source electron incident on the plane of the corrector at $z_{\rm PAC}$, while $\mathcal{T}_{\rm PAC}$ and $\mathcal{T}_{\rm L}$ account for transmission through the PAC and electron lenses (orange and blue frames in Fig.~\ref{Fig1}, respectively). The latter is for a thin lens well described by \cite{AOP01,PPB18}
\begin{align}
\mathcal{T}_{\rm L}\big{|}_{RR'}=\delta(R-R')\,\ee^{\ii\chi(\theta)}\,\ee^{-\ii q_0 R^2/2f}\,\Theta(R_{\rm max}-R),
\label{TL}
\end{align}
where $f$ is the focal distance, a pupil blocks propagation above a radial distance $R_{\rm max}$, and the phase $\chi(\theta)$ accounts for aberrations in the lenses as a function of exit angle $\theta=R/(z_{\rm sample}-z_{\rm L,out})$. Here, we concentrate on spherical aberration, so we express $\chi(\theta)=C_3 q_0 \theta^4/4$ in terms of the (length) coefficient $C_3$ \cite{AOP01,PPB18}. For simplicity, in what follows we consider a point source producing a spherical wave $\psi_{z_{\rm PAC}}(R)\propto\ee^{\ii q_0R^2/2(z_{\rm PAC}-z_{\rm source})}$ with $m=0$. Additionally, we take the PAC to coincide with the near and far sides of the optical lenses at the virtual plane $z=z_L$.

\begin{figure*}[ht]
\centering \includegraphics[width=\textwidth]{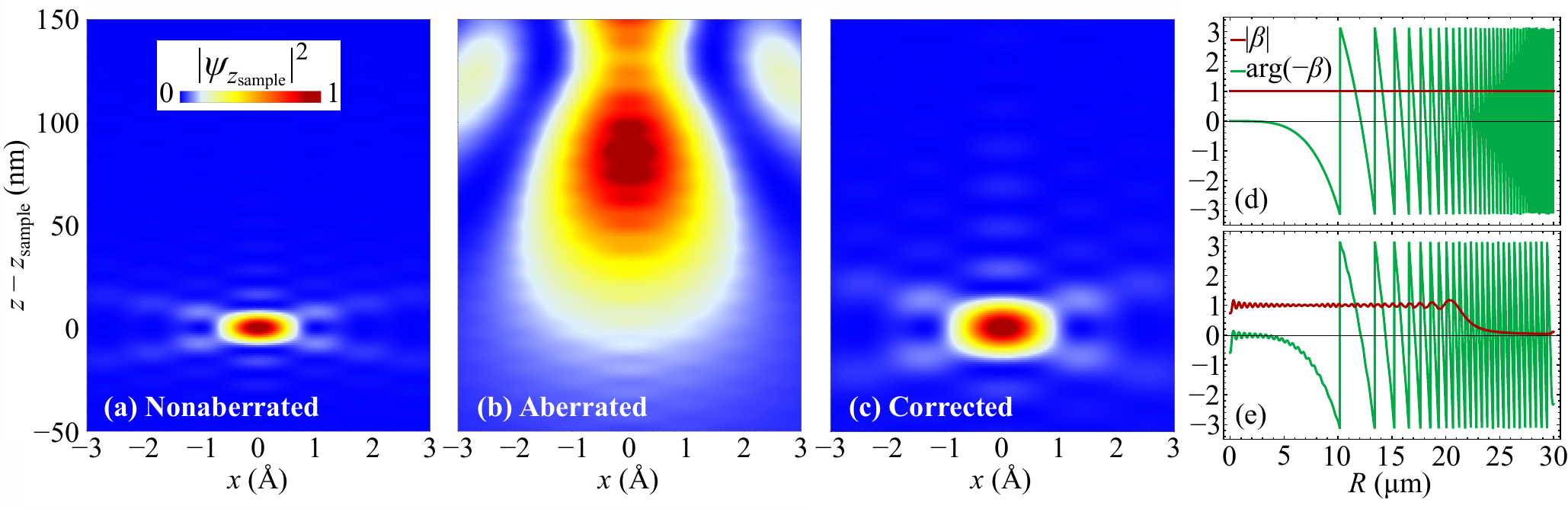}
\caption{Spatial dependence of the electron intensity focus at the sample. We plot the normalized beam electron density $\abs{\psi_{\rm sample}}^2$ [Eq.~\eqref{micros}] obtained through (a) aberration-free electron optics; (b) electron optics introducing spherical aberration with $C_3=1\,$mm; and (c) same as (b) including a PAC module. We consider 60-keV electrons, a focal distance $f=1\,$mm, $z_{\rm L}-z_{\rm source}=40f$, and an aperture $R_\mathrm{max}=30\,\upmu\mathrm{m}$. The corrected beam profile in (c) is calculated by using a realistic spatial dependence of the coupling parameter $\beta$ as a function of radial distance $R$ in the PAC [panel (e), obtained from Eqs.~\eqref{betarealistic} and \eqref{betakpar} with $\lambda_0=500\,$nm light wavelength], which differs from the ideal $\beta$ that is needed to perfectly correct the aberration [panel (d), Eq.~\eqref{argbeta}].} 
\label{Fig2}
\end{figure*}


In the absence of aberrations ($\chi=0$), $\psi_{z_{\rm sample}}$ is focused at a position $z=z_{\rm sample}$ determined by the lens formula $1/(z_{\rm sample}-z_{\rm L})+1/(z_{\rm L}-z_{\rm source})=1/f$ (see Appendix\ \ref{AS2}). This is illustrated in Fig.~\ref{Fig2}(a) for 60-keV electrons ($\sim5\,$pm wavelength) with $z_{\rm L}-z_{\rm source}=40f$ (implying $z_{\rm source}-z_{\rm L}\approx f$), $f=1\,$mm, and $R_\mathrm{max}=30\,\upmu$m. The focal spot is limited by diffraction at the aperture, yielding a $\psi_{z_{\rm sample}}\sim J_1(R/\Delta)$ transversal profile of sub-{\AA}ngstrom width $\sim\Delta=f/q_0R_\mathrm{max}=0.26\,${\AA}. In contrast, a typical spherical aberration corresponding to $C_3=1\,$mm produces a substantially broadened and shifted focus [Fig.~\ref{Fig2}(b)], accompanied by satellite foci along the optical axis.



\section{Electron optical phase plate}

We intend to cancel the aberration phase $\chi$ by imprinting an additional phase on the electron wave function through the interaction with an optical near field \cite{paper272,paper311,paper312}. For this purpose, we consider an optically-opaque electron-transparent film subject to external illumination, a configuration that has been demonstrated to produce large coupling to the electrons \cite{paper311}. The transmitted electron wave function consists of sidebands of energies $\mathcal{E}_0+\hbar\ell\omega_0$ separated from the incident energy by multiples $\ell$ ($<0$ for loss, $>0$ for gain) of the photon energy $\hbar\omega_0$. In the nonrecoil approximation, the wave function associated with each transmitted sideband $\ell$ consists of the incident wave function times a multiplicative factor accounted for by the operator \cite{paper311,paper339}
\begin{align}
\mathcal{T}_{\rm PAC}\big{|}_{RR'}=\delta(R-R')\,J_\ell(2\abs{\beta})\,\ee^{\ii\ell\,\mathrm{arg}\left\lbrace-\beta\right\rbrace},
\label{TPAC}
\end{align}
where the coupling coefficient
\begin{align}
\beta(\Rb)=\frac{e}{\hbar\omega_0}\int_{-\infty}^\infty dz\,E_z(\Rb,z)\,\ee^{-\ii \omega_0 z/v}
\label{beta}
\end{align}
captures the electron-light interaction through the (along-the-beam) $E_z$ component of the optical electric-field amplitude, which bears a dependence on transversal coordinates $\Rb$ that can be controlled through a spatial light modulator. We first consider axially-symmetric illumination [implying $m=0$ in Eq.~\eqref{micros}] and express the incident optical field $E_z=\int_0^{k_0}d\kpar\,J_0(\kpar R)\,\ee^{\ii k_z z}\,\alpha_{\kpar}$ as a combination of cylindrical Bessel waves with in- and out-of-plane wave vector components $\kpar$ and $k_z=\sqrt{k_0^2-\kpar^2}$, respectively, limited by the free-space light wave vector $k_0=\omega_0/c$. Upon insertion of this field in Eq.~\eqref{beta}, we find
\begin{align}
\beta(R)=\int_0^{k_0} \kpar\,d\kpar\,J_0(\kpar R)\,\beta_{\kpar},
\label{betarealistic}
\end{align}
where the coefficient $\beta_{\kpar}$ is proportional to $\alpha_{\kpar}$ and also includes light components reflected at the film \cite{paper311}; we stress that $\beta_{\kpar}$ can therefore be controlled through the applied angular light profile $\alpha_{\kpar}$.



\section{Design of the PAC field profile}

\begin{figure}[ht]
\centering
\includegraphics[width=0.45\textwidth]{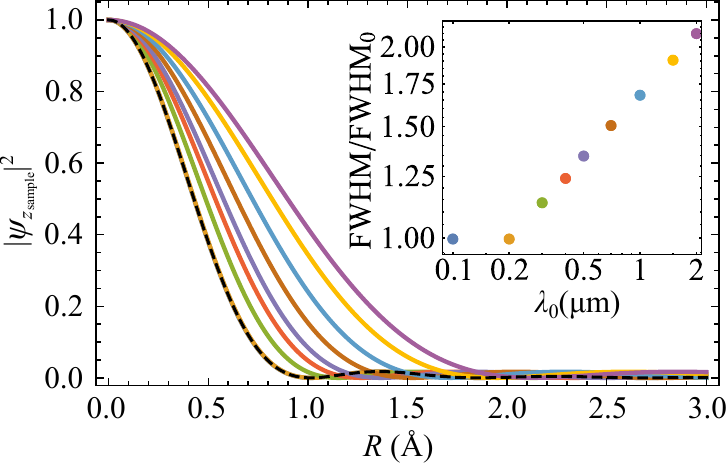}
\caption{Approaching perfect aberration correction. We show the focal spot profile obtained with the same parameters as in Fig.~\ref{Fig2}(c) using light of different wavelength $\lambda_0$. The inset shows the spot FWHM normalized to the unaberrated one (FWHM$_0$, corresponding to the dashed black curve in the main plot) and provides a color legend for the main plot.}
\label{Fig3}
\end{figure}

We now design the PAC based on an electron optical phase plate in which $\ell=1$ is selected, while other sidebands ($\ell\neq1$) are filtered out by a monochromator (Fig.~\ref{Fig1}), using for example a Wien filter \cite{HS19}. The aberration phase $\chi$ introduced through $\mathcal{T}_{\rm L}$ [Eq.~\eqref{TL}] can then be eliminated from Eq.~\eqref{micros} by setting
\begin{align}
\mathrm{arg}\left\lbrace-\beta(R)\right\rbrace=-\chi(\theta)
\label{argbeta}
\end{align}
in $\mathcal{T}_{\rm PAC}$ [Eq.~\eqref{TL}], where $R=\theta(z_{\rm sample}-z_{\rm L})$. We can maximize the current by imposing $|\beta|=\beta_0\approx0.92$, which yields an absolute maximum fraction of the $\ell=1$ sideband $J_1^2(2\beta_0)\approx34$\% (see Appendix\ \ref{AS3}). The PAC then involves a reduction in electron current by a factor of $\sim1/3$. The spatial dependence of $\beta=-\beta_0\ee^{-\ii\chi}$ required to produce perfect aberration correction and maximum $\ell=1$ current is presented in Fig.~\ref{Fig2}(d) according to Eq.~\eqref{argbeta} for $\mathcal{E}_0=60\,$keV and $f=C_3=1\,$mm. However, optical diffraction at the used finite light wavelength $\lambda_0=2\pi/k_0$ limits the profile of $\beta$ that can be achieved in practice using far-field illumination. We find a nearly optimum realistic profile by setting $\beta=-\beta_0\ee^{-\ii\chi}$ in Eq.~\eqref{betarealistic} and approximately inverting this equation to yield
\begin{align}
    \beta_{k_\parallel}=-\beta_0\int_0^{R_{\rm max}}\!\!\!\!\!R\,dR\,J_0(\kpar R)\,\ee^{-\ii \chi[R/(z_{\rm sample}-z_{\rm L})]},
    \label{betakpar}
\end{align}
(this inversion is only exact in the $k_0R_{\rm max}\gg1$ limit). The coupling coefficient obtained by inserting Eq.~\eqref{betakpar} back into Eq.~\eqref{betarealistic} is plotted in Fig.~\ref{Fig2}(e) for a photon wavelength $\lambda_0=500\,$nm, which resembles the perfect-correction coefficient of Fig.~\ref{Fig2}(d) up to $R\sim20\,\upmu$m, but deviates substantially from that target value at larger $R$ (i.e., where the phase $\chi$ exhibits rapid variations over a distance $\sim\lambda_0$). Although the resulting corrected beam profile plotted in Fig.~\ref{Fig2}(c) is not perfect, it still provides an impressive improvement in electron focusing compared to the aberrated spot shown in Fig.~\ref{Fig2}(b) (see Appendix\ \ref{AS3}). We note that the degree of correction increases when $\lambda_0$ is made smaller relative to $R_{\rm max}$, as we show in Fig.~\ref{Fig3}, which further predicts a remakable aberration compensation using blue light.

\begin{figure}[ht]
\centering
\includegraphics[width=0.45\textwidth]{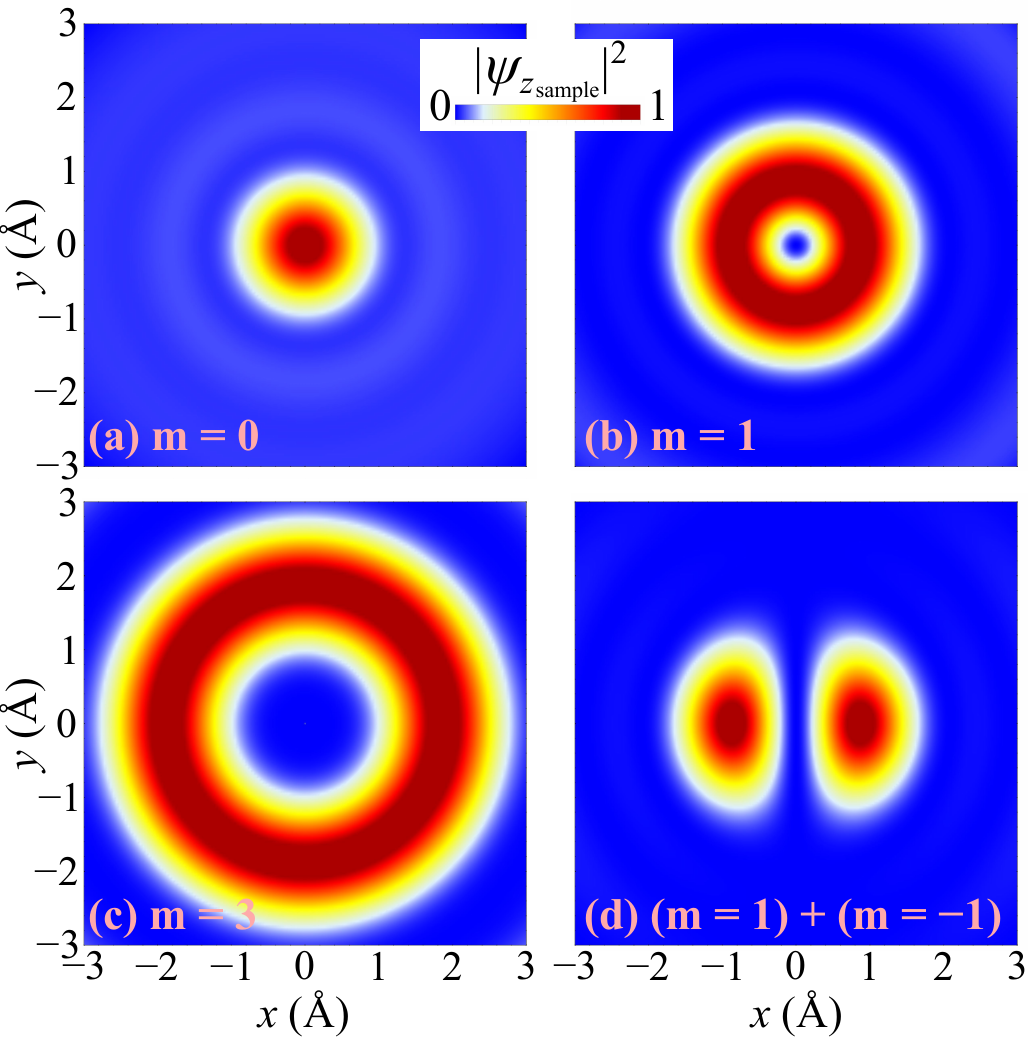}
\caption{Manipulation of the focal spot profile. Transversal cuts of electron focal spots obtained by correcting spherical aberration under the same conditions as in Fig.~\ref{Fig2}(c) but using light fields with the indicated symmetry: $\ee^{\ii m\varphi}$ azimuthal dependence in (a-c) and $\cos^2\varphi$ in (d).}
\label{Fig4}
\end{figure}


The PAC can be fed using light with definite chirality $m$ [i.e., $E_z(\Rb,z)=E_z(R,z)\ee^{\ii m\varphi_\Rb}$, which directly translates through Eqs.~\eqref{TPAC} and \eqref{beta} into $\mathcal{T}_{\rm PAC}\propto\ee^{\ii m\varphi_\Rb}$], still described by Eqs.~\eqref{betarealistic} and \eqref{betakpar} with $J_0$ substituted by $J_m$. Results for the transversal profile of the electron focus using chiral PACs with $m=1$ and 3 are compared with the $m=0$ profile in Fig.~\ref{Fig4}, revealing the formation of donuts associated with an electron wave function of $\ee^{\ii m\varphi}$ azimuthal symmetry. More complex profiles are possible, which should be reachable using a spatial light modulator to project light on the PAC. For example, Fig.~\ref{Fig4}(d) shows the result for a plot obtained by projecting a symmetric combination of $m=1$ and $m=-1$ light.


\section{Conclusion}

In summary, we propose the use of electron optical phase plates as a way of tailoring the amplitude and phase of the electron transversal wave function in an electron beam. Specifically, we theoretically demonstrate correction of spherical aberration without the involvement of complex electron-optics elements. This concept can be straightforwardly applied to eliminate any undesired distortions introduced by electron optics in both standard and ultrafast electron microscopes. Although we apply analytical methods to produce a proof-of-principle design, substantial improvement in the performance of such phase plates could be gained through machine learning, which should enable the design of more complex electron spot shapes. Ultimately, iterative improvement of the PAC could be attained through a feedback loop involving measurement of the electron spot and modification of the projected light profile. Additionally, temporal manipulation of the imprinted optical phase offers interesting possibilities for the exploration of sample dynamics through time-varying electron spot profiles. The versatility and compactness of electron optical phase plates hold potential for active control of electron wave functions beyond the present application in aberration correction.

\acknowledgments

We thank Mathieu Kociak, Albert Polman, and Claus Ropers for helpful and enjoyable discussions. This work has been supported in part by the Spanish MINECO (MAT2017-88492-R and SEV2015-0522), ERC (Advanced Grant 789104-eNANO), the Catalan CERCA Program, and Fundaci\'{o} Privada Cellex.

\appendix

\begin{widetext}

\section{Derivation of Eq. (1) in the main text}
\label{AS1}

Because a monochromatic electron wave function $\psi_z(\Rb)$ satisfies the Helmholz equation $(\nabla^2_\Rb+\partial_{zz}-q_0^2)\psi_z(\Rb)=0$ for fixed electron wave number $q_0$, its propagation from a transversal plane $z'$ to $z$ can be realized through the integral
\begin{align}
\psi_z(\Rb)=\int\frac{d^2\Qb}{(2\pi)^2}\,\ee^{\ii[\Qb\cdot\Rb+q_z(z-z')]}
\int d^2\Rb'\ee^{-\ii\Qb\cdot\Rb'}\psi_{z'}(\Rb'),\label{propa1}
\end{align}
where we use the notation $\Rb=(x,y)$, the $\Rb'$ integral produces the Fourier transform along the transversal directions, each component of wave vector $(\Qb,q_z)$ with $q_z=\sqrt{q_0^2-Q^2+\ii0^+}$ and ${\rm Im}\{q_z\}>0$ evolves as a plane wave, and the $\Qb$ integral yields the inverse Fourier transform after propagation. We note that evanescent waves with $Q>q_0$ die away in the $q_0(z-z')\gg1$ limit, so only $Q<q_0$ contributes to the integral in Eq.\ (\ref{propa1}) for propagation along a distance $z-z'$ spanning many electron wavelengths. Additionally, in the paraxial approximation for well collimated beams, only small components $Q\ll q_0$ contribute to Eq.\ (\ref{propa1}), so we can approximate $q_z\approx q_0-Q^2/2q_0$. Using this expression and considering wave functions with an azimuthal dependence given by $\psi_z(\Rb)=\psi_z(R)\ee^{\ii m\varphi_\Rb}$ in terms of a well-defined angular momentum number $m$, we can perform the integrals over $\varphi_{\Rb'}$ and $\varphi_\Qb$ in Eq.\ (\ref{propa1}) using the identity (Eq.\ (9.1.21) of Ref. \citenum{AS1972}) $\int_0^{2\pi}d\varphi\,\ee^{\pm\ii z\cos\varphi}\ee^{\ii m\varphi}=2\pi \ii^{\pm m}J_m(z)$. We find
\begin{align}
\psi_z(R)&=\ee^{\ii q_0(z-z')}\int_0^\infty QdQ\,J_m(Q R)\,\ee^{-\ii Q^2(z-z')/2q_0}
\int_0^\infty R'dR'\,J_m(Q R')\,\psi_{z'}(R')\nonumber\\
&=\frac{(-\ii)^{m+1}q_0}{z-z'}\,\ee^{\ii q_0(z-z')}\int R'dR'\,\exp\left[\frac{\ii q_0(R^2+R'^2)}{2(z-z')}\right]\,J_m\left(\frac{q_0RR'}{z-z'}\right)\,\psi_{z'}(R'),
\nonumber
\end{align}
where the second line, which coincides with Eq.\ (1) in the main text, is obtained after exchanging the order of integration and applying the Weber second exponential integral (Sec.\ 13.31 of Ref. \citenum{W1944}) $\int_0^\infty \theta d\theta\,J_m(\theta x)J_m(\theta x')\,\ee^{-\ii\theta^2/2}=(-\ii)^{m+1}\,\ee^{\ii(x^2+x'^2)/2}J_m(xx')$.

\section{Calculation of focal spots}
\label{AS2}

For simplicity, we assume the near and far sides of the electron-optics lens system to coincide with the PAC in a virtual plane at $z_{\rm L}$. In practice, this allows us to set $z_{\rm PAC}=z_{\rm L,in}=z_{\rm L,out}=z_{\rm L}$ in Eq.\ (2), which, noticing that $\mathcal{T}^m_{\Delta z}$ becomes the identity operator when $\Delta z=0$, reduces to $\psi_{z_{\rm sample}}=\mathcal{F}^m_{f'}\cdot \mathcal{T}_{\rm L}\cdot \mathcal{T}_{\rm PAC}\cdot \psi_{z_{\rm PAC}}$, where $f'=z_{\rm sample}-z_{\rm L}$. Inserting in this equation the definitions provided in the main text for the different factors and considering illumination of the PAC with a chiral light field $E_z(\Rb,z)\propto\ee^{\ii m\varphi_\Rb}$, we find the expression
\begin{align}
\psi_z(\Rb)&\propto\ee^{\ii[m\varphi_\Rb+q_0z+q_0R^2/2f']}
\int_0^{R_{\rm max}} \!\!R'dR' \;J_m\left(\frac{q_0RR'}{z-z_{\rm L}}\right) \;J_1(2\abs{\beta(R')})
\nonumber\\
&\times \ee^{\ii\chi(R'/f')+\ii\,{\rm arg}\{-\beta(R')\}}\;
\ee^{\ii(q_0R'^2/2)[1/(z-z_{\rm L})+1/(z_{\rm L}-z_{\rm source})-1/f]}
\nonumber
\end{align}
for the electron wave function in the sample region, where $\chi(R/f')=C_3q_0R^4/4f'^4$ and we consider points separated by a small distance from the focal point $(\Rb,z)=(0,z_{\rm sample})$ compared with $f'$, where $z_{\rm sample}$ is defined by the condition that the argument of the last exponential vanishes, leading to the lens equation $1/(z_{\rm sample}-z_{\rm L})+1/(z_{\rm L}-z_{\rm source})=1/f$.

\section{Additional figures}
\label{AS3}

We show in Fig.\ \ref{FigS1} further details of the electron optical phase plate used for the PAC in the main text. In Fig.\ \ref{FigS2} we present longitudinal and transversal cuts of the spots plotted in Fig.\ 2 of the main text to facilitate their comparison.

\begin{figure}[ht]
\centering
\includegraphics[width=0.75\textwidth]{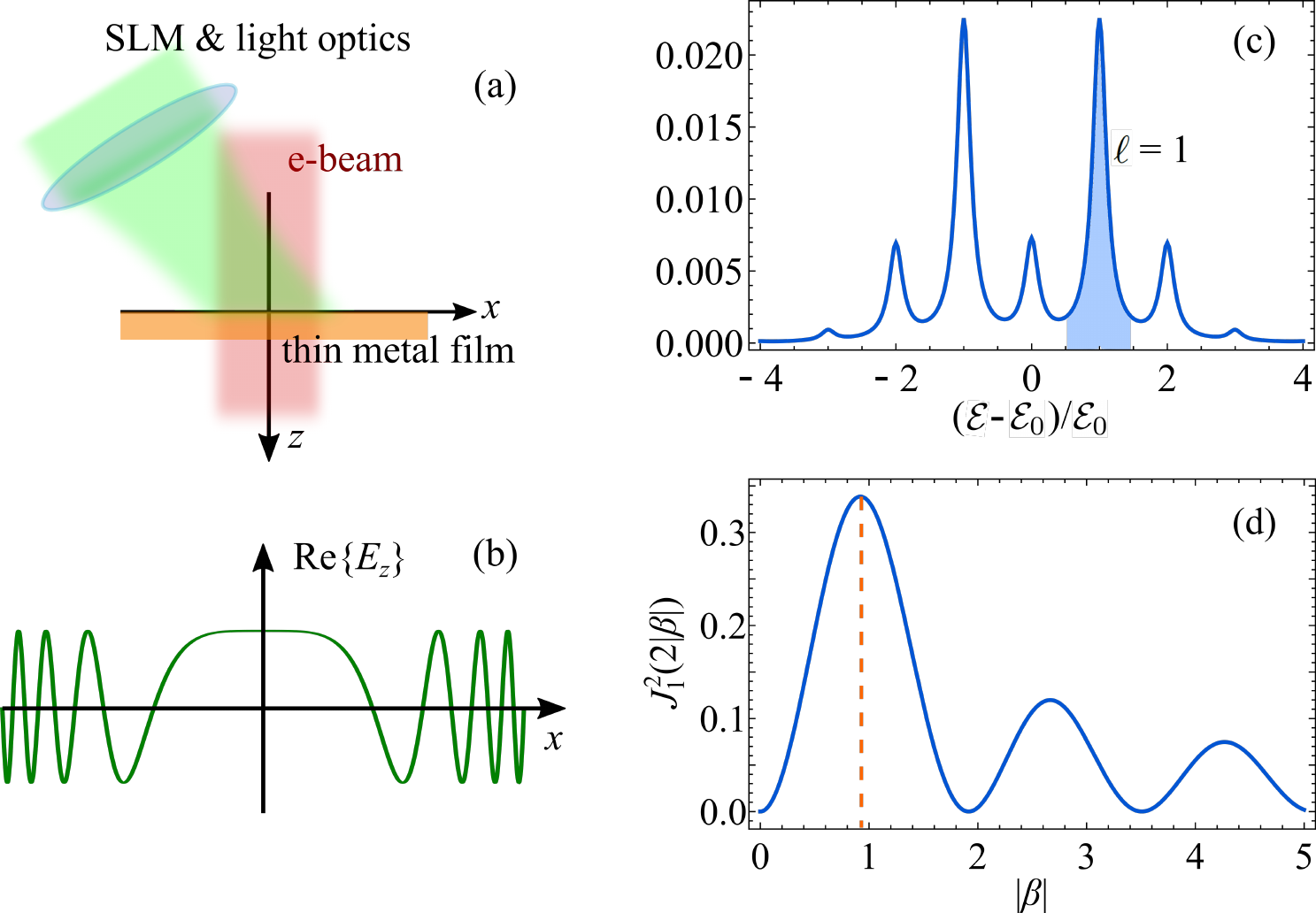}
\caption{Details of the electron optical phase plate used for the PAC element proposed in this paper. (a) Light is projected on a thin film with a spatial distribution of phase and amplitude delineated through an spatial light modulator (SLM). The film is taken to be nearly transparent to the electrons, but totally reflective to the light. (b) An optical electric field is then established as a function of lateral position along the film, resulting from the sum of incident and reflected light. Only the out-of-plane electric field $E_z$ couples to the electron moving along $z$. (c) Interaction with $E_z$ produces inelastic sidebands in the spectrum of the transmitted electrons, separated by multiples of the photon energy $\hbar\omega_0$ with respect to the incident electron energy $\mathcal{E}_0$. (d) A monochromator is used to retain electrons only within the first sideband $\ell=1$. The transmitted wave function consists of the incident wave function times a factor $J_1(2|\beta|)\ee^{\ii{\rm arg}\{-\beta\}}$, where the coupling parameter $\beta$ scales linearly with $E_z$ and therefore varies with lateral position along the film. The field amplitude is adjusted to take values in the region to the left of the vertical orange line [$|\beta|<\beta_0\approx0.92$ below the maximum first-sideband transmission intensity $J_1^2(2\beta_0)\approx34$\%].}
\label{FigS1}
\end{figure}

\begin{figure}[ht]
\centering
\includegraphics[width=0.85\textwidth]{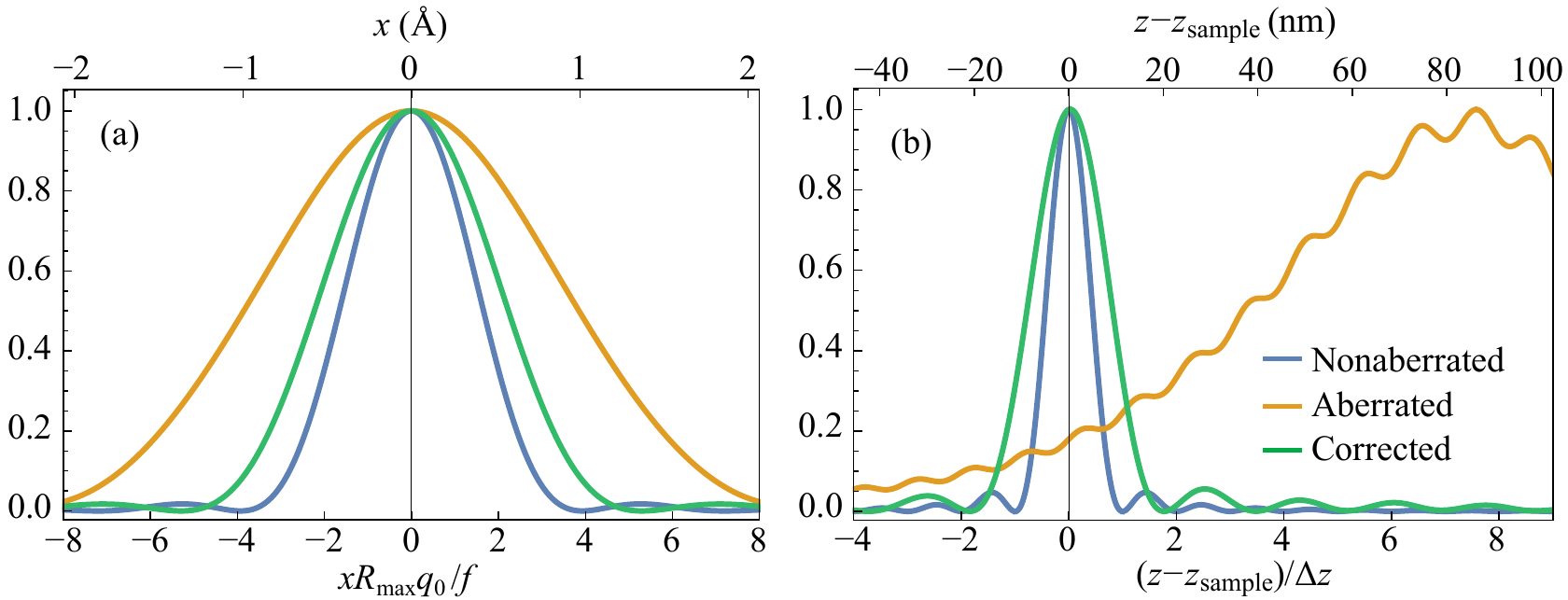}
\caption{Profiles showing the electron probability $\abs{\psi_{z_{\rm sample}}}^2$ along transversal (a) and longitudinal (b) cuts across the focal spot maximum for a nonaberrated beam and aberrated beams under the conditions of Fig.\ 2 in the main text. The profiles are normalized to their respective maxima. The lower $x$ scale in (a) is normalized such that the unaberrated curve is universal. The lower $z$ scale in (b) is normalized using $\Delta z=1/[4\pi/(q_0 R_\mathrm{max}^2)+1/f-1/(z_{\rm sample}-z_{\rm L,out})]-1/[1/f-1/(z_{\rm sample}-z_{\rm L,out})]$. The upper horizontal scales in both (a) and (b) are obtained for $R_\mathrm{max}=30\,\upmu\mathrm{m}$.} 
\label{FigS2}
\end{figure}

\end{widetext}


\begin{thebibliography}{53}
\expandafter\ifx\csname natexlab\endcsname\relax\def\natexlab#1{#1}\fi
\expandafter\ifx\csname bibnamefont\endcsname\relax
  \def\bibnamefont#1{#1}\fi
\expandafter\ifx\csname bibfnamefont\endcsname\relax
  \def\bibfnamefont#1{#1}\fi
\expandafter\ifx\csname citenamefont\endcsname\relax
  \def\citenamefont#1{#1}\fi
\expandafter\ifx\csname url\endcsname\relax
  \def\url#1{\texttt{#1}}\fi
\expandafter\ifx\csname urlprefix\endcsname\relax\def\urlprefix{URL }\fi
\providecommand{\bibinfo}[2]{#2}
\providecommand{\eprint}[2][]{\url{#2}}

\bibitem[{\citenamefont{Hawkes and Spence}(2019)}]{HS19}
\bibinfo{author}{\bibfnamefont{P.}~\bibnamefont{Hawkes}} \bibnamefont{and}
  \bibinfo{author}{\bibfnamefont{J.}~\bibnamefont{Spence}},
  \emph{\bibinfo{title}{Springer Handbook of Microscopy}}
  (\bibinfo{publisher}{Springer Nature Switzerland AG}, \bibinfo{year}{2019}).

\bibitem[{\citenamefont{Batson et~al.}(2002)\citenamefont{Batson, Dellby, and
  Krivanek}}]{BDK02}
\bibinfo{author}{\bibfnamefont{P.~E.} \bibnamefont{Batson}},
  \bibinfo{author}{\bibfnamefont{N.}~\bibnamefont{Dellby}}, \bibnamefont{and}
  \bibinfo{author}{\bibfnamefont{O.~L.} \bibnamefont{Krivanek}},
  \bibinfo{journal}{Nature} \textbf{\bibinfo{volume}{418}},
  \bibinfo{pages}{617} (\bibinfo{year}{2002}).

\bibitem[{\citenamefont{Muller et~al.}(2008)\citenamefont{Muller, {Fitting
  Kourkoutis}, Murfitt, Song, Hwang, Silcox, Dellby, and Krivanek}}]{MKM08}
\bibinfo{author}{\bibfnamefont{D.~A.} \bibnamefont{Muller}},
  \bibinfo{author}{\bibfnamefont{L.}~\bibnamefont{{Fitting Kourkoutis}}},
  \bibinfo{author}{\bibfnamefont{M.}~\bibnamefont{Murfitt}},
  \bibinfo{author}{\bibfnamefont{J.~H.} \bibnamefont{Song}},
  \bibinfo{author}{\bibfnamefont{H.~Y.} \bibnamefont{Hwang}},
  \bibinfo{author}{\bibfnamefont{J.}~\bibnamefont{Silcox}},
  \bibinfo{author}{\bibfnamefont{N.}~\bibnamefont{Dellby}}, \bibnamefont{and}
  \bibinfo{author}{\bibfnamefont{O.~L.} \bibnamefont{Krivanek}},
  \bibinfo{journal}{Science} \textbf{\bibinfo{volume}{319}},
  \bibinfo{pages}{1073} (\bibinfo{year}{2008}).

\bibitem[{\citenamefont{Jiang et~al.}(2018)\citenamefont{Jiang, Chen, Han, Deb,
  Gao, Xie, Purohit, Tate, Park, Gruner et~al.}}]{JCH18}
\bibinfo{author}{\bibfnamefont{Y.}~\bibnamefont{Jiang}},
  \bibinfo{author}{\bibfnamefont{Z.}~\bibnamefont{Chen}},
  \bibinfo{author}{\bibfnamefont{Y.}~\bibnamefont{Han}},
  \bibinfo{author}{\bibfnamefont{P.}~\bibnamefont{Deb}},
  \bibinfo{author}{\bibfnamefont{H.}~\bibnamefont{Gao}},
  \bibinfo{author}{\bibfnamefont{S.}~\bibnamefont{Xie}},
  \bibinfo{author}{\bibfnamefont{P.}~\bibnamefont{Purohit}},
  \bibinfo{author}{\bibfnamefont{M.~W.} \bibnamefont{Tate}},
  \bibinfo{author}{\bibfnamefont{J.}~\bibnamefont{Park}},
  \bibinfo{author}{\bibfnamefont{S.~M.} \bibnamefont{Gruner}},
  \bibnamefont{et~al.}, \bibinfo{journal}{Nature}
  \textbf{\bibinfo{volume}{559}}, \bibinfo{pages}{343} (\bibinfo{year}{2018}).

\bibitem[{\citenamefont{Midgley and Dunin-Borkowski}(2009)}]{MD09}
\bibinfo{author}{\bibfnamefont{P.~A.} \bibnamefont{Midgley}} \bibnamefont{and}
  \bibinfo{author}{\bibfnamefont{R.~E.} \bibnamefont{Dunin-Borkowski}},
  \bibinfo{journal}{Nat.\ Mater.} \textbf{\bibinfo{volume}{8}},
  \bibinfo{pages}{271} (\bibinfo{year}{2009}).

\bibitem[{\citenamefont{Fink et~al.}(1997)\citenamefont{Fink, Schmid,
  Ermantraut, and Schulz}}]{FSE97}
\bibinfo{author}{\bibfnamefont{H.-W.} \bibnamefont{Fink}},
  \bibinfo{author}{\bibfnamefont{H.}~\bibnamefont{Schmid}},
  \bibinfo{author}{\bibfnamefont{E.}~\bibnamefont{Ermantraut}},
  \bibnamefont{and} \bibinfo{author}{\bibfnamefont{T.}~\bibnamefont{Schulz}},
  \bibinfo{journal}{J.\ Opt.\ Soc.\ Am.\ A} \textbf{\bibinfo{volume}{14}},
  \bibinfo{pages}{2168} (\bibinfo{year}{1997}).

\bibitem[{\citenamefont{Yasin et~al.}(2018)\citenamefont{Yasin, Harvey, Chess,
  Pierce, Ophus, Ercius, and McMorran}}]{YHC18}
\bibinfo{author}{\bibfnamefont{F.~S.} \bibnamefont{Yasin}},
  \bibinfo{author}{\bibfnamefont{T.~R.} \bibnamefont{Harvey}},
  \bibinfo{author}{\bibfnamefont{J.~J.} \bibnamefont{Chess}},
  \bibinfo{author}{\bibfnamefont{J.~S.} \bibnamefont{Pierce}},
  \bibinfo{author}{\bibfnamefont{C.}~\bibnamefont{Ophus}},
  \bibinfo{author}{\bibfnamefont{P.}~\bibnamefont{Ercius}}, \bibnamefont{and}
  \bibinfo{author}{\bibfnamefont{B.~J.} \bibnamefont{McMorran}},
  \bibinfo{journal}{Nano\ Lett.} \textbf{\bibinfo{volume}{18}},
  \bibinfo{pages}{7188} (\bibinfo{year}{2018}).

\bibitem[{\citenamefont{McCartney et~al.}(2010)\citenamefont{McCartney,
  Agarwal, Chung, Cullen, Han, He, Li, Wang, Zhou, and Smith}}]{MAC10}
\bibinfo{author}{\bibfnamefont{M.~R.} \bibnamefont{McCartney}},
  \bibinfo{author}{\bibfnamefont{N.}~\bibnamefont{Agarwal}},
  \bibinfo{author}{\bibfnamefont{S.}~\bibnamefont{Chung}},
  \bibinfo{author}{\bibfnamefont{D.~A.} \bibnamefont{Cullen}},
  \bibinfo{author}{\bibfnamefont{M.-G.} \bibnamefont{Han}},
  \bibinfo{author}{\bibfnamefont{K.}~\bibnamefont{He}},
  \bibinfo{author}{\bibfnamefont{L.}~\bibnamefont{Li}},
  \bibinfo{author}{\bibfnamefont{H.}~\bibnamefont{Wang}},
  \bibinfo{author}{\bibfnamefont{L.}~\bibnamefont{Zhou}}, \bibnamefont{and}
  \bibinfo{author}{\bibfnamefont{D.~J.} \bibnamefont{Smith}},
  \bibinfo{journal}{Ultramicroscopy} \textbf{\bibinfo{volume}{110}},
  \bibinfo{pages}{375} (\bibinfo{year}{2010}).

\bibitem[{\citenamefont{Nicoletti et~al.}(2013)\citenamefont{Nicoletti,
  de~La~Pe{\~n}a, Leary, Holland, Ducati, and Midgley}}]{NPL13}
\bibinfo{author}{\bibfnamefont{O.}~\bibnamefont{Nicoletti}},
  \bibinfo{author}{\bibfnamefont{F.}~\bibnamefont{de~La~Pe{\~n}a}},
  \bibinfo{author}{\bibfnamefont{R.~K.} \bibnamefont{Leary}},
  \bibinfo{author}{\bibfnamefont{D.~J.} \bibnamefont{Holland}},
  \bibinfo{author}{\bibfnamefont{C.}~\bibnamefont{Ducati}}, \bibnamefont{and}
  \bibinfo{author}{\bibfnamefont{P.~A.} \bibnamefont{Midgley}},
  \bibinfo{journal}{Nature} \textbf{\bibinfo{volume}{502}}, \bibinfo{pages}{80}
  (\bibinfo{year}{2013}).

\bibitem[{\citenamefont{Shibata et~al.}(2017)\citenamefont{Shibata, Kov\'acs,
  Kiselev, Kanazawa, Dunin-Borkowski, and Tokura}}]{SKK17}
\bibinfo{author}{\bibfnamefont{K.}~\bibnamefont{Shibata}},
  \bibinfo{author}{\bibfnamefont{A.}~\bibnamefont{Kov\'acs}},
  \bibinfo{author}{\bibfnamefont{N.~S.} \bibnamefont{Kiselev}},
  \bibinfo{author}{\bibfnamefont{N.}~\bibnamefont{Kanazawa}},
  \bibinfo{author}{\bibfnamefont{R.~E.} \bibnamefont{Dunin-Borkowski}},
  \bibnamefont{and} \bibinfo{author}{\bibfnamefont{Y.}~\bibnamefont{Tokura}},
  \bibinfo{journal}{Phys.\ Rev.\ Lett.} \textbf{\bibinfo{volume}{118}},
  \bibinfo{pages}{087202} (\bibinfo{year}{2017}).

\bibitem[{\citenamefont{Guzzinati et~al.}(2017)\citenamefont{Guzzinati, Beche,
  Lourenco-Martins, Martin, Kociak, and Verbeeck}}]{GBL17}
\bibinfo{author}{\bibfnamefont{G.}~\bibnamefont{Guzzinati}},
  \bibinfo{author}{\bibfnamefont{A.}~\bibnamefont{Beche}},
  \bibinfo{author}{\bibfnamefont{H.}~\bibnamefont{Lourenco-Martins}},
  \bibinfo{author}{\bibfnamefont{J.}~\bibnamefont{Martin}},
  \bibinfo{author}{\bibfnamefont{M.}~\bibnamefont{Kociak}}, \bibnamefont{and}
  \bibinfo{author}{\bibfnamefont{J.}~\bibnamefont{Verbeeck}},
  \bibinfo{journal}{Nat.\ Commun.} \textbf{\bibinfo{volume}{8}},
  \bibinfo{pages}{14999} (\bibinfo{year}{2017}).

\bibitem[{\citenamefont{Lloyd et~al.}(2017)\citenamefont{Lloyd, Babiker,
  Thirunavukkarasu, and Yuan}}]{LBT17}
\bibinfo{author}{\bibfnamefont{S.~M.} \bibnamefont{Lloyd}},
  \bibinfo{author}{\bibfnamefont{M.}~\bibnamefont{Babiker}},
  \bibinfo{author}{\bibfnamefont{G.}~\bibnamefont{Thirunavukkarasu}},
  \bibnamefont{and} \bibinfo{author}{\bibfnamefont{J.}~\bibnamefont{Yuan}},
  \bibinfo{journal}{Rev.\ Mod.\ Phys.} \textbf{\bibinfo{volume}{89}},
  \bibinfo{pages}{035004} (\bibinfo{year}{2017}).

\bibitem[{\citenamefont{Rusz and Bhowmick}(2013)}]{RB13_2}
\bibinfo{author}{\bibfnamefont{J.}~\bibnamefont{Rusz}} \bibnamefont{and}
  \bibinfo{author}{\bibfnamefont{S.}~\bibnamefont{Bhowmick}},
  \bibinfo{journal}{Phys.\ Rev.\ Lett.} \textbf{\bibinfo{volume}{111}},
  \bibinfo{pages}{105504} (\bibinfo{year}{2013}).

\bibitem[{\citenamefont{Verbeeck et~al.}(2010)\citenamefont{Verbeeck, Tian, and
  Schattschneider}}]{VTS10}
\bibinfo{author}{\bibfnamefont{J.}~\bibnamefont{Verbeeck}},
  \bibinfo{author}{\bibfnamefont{H.}~\bibnamefont{Tian}}, \bibnamefont{and}
  \bibinfo{author}{\bibfnamefont{P.}~\bibnamefont{Schattschneider}},
  \bibinfo{journal}{Nature} \textbf{\bibinfo{volume}{467}},
  \bibinfo{pages}{301} (\bibinfo{year}{2010}).

\bibitem[{\citenamefont{McMorran et~al.}(2011)\citenamefont{McMorran, Agrawal,
  Anderson, Herzing, Lezec, McClelland, and Unguris}}]{MAA11}
\bibinfo{author}{\bibfnamefont{B.~J.} \bibnamefont{McMorran}},
  \bibinfo{author}{\bibfnamefont{A.}~\bibnamefont{Agrawal}},
  \bibinfo{author}{\bibfnamefont{I.~M.} \bibnamefont{Anderson}},
  \bibinfo{author}{\bibfnamefont{A.~A.} \bibnamefont{Herzing}},
  \bibinfo{author}{\bibfnamefont{H.~J.} \bibnamefont{Lezec}},
  \bibinfo{author}{\bibfnamefont{J.~J.} \bibnamefont{McClelland}},
  \bibnamefont{and} \bibinfo{author}{\bibfnamefont{J.}~\bibnamefont{Unguris}},
  \bibinfo{journal}{Science} \textbf{\bibinfo{volume}{331}},
  \bibinfo{pages}{192} (\bibinfo{year}{2011}).

\bibitem[{\citenamefont{Shiloh et~al.}(2014)\citenamefont{Shiloh, Lereah,
  Lilach, and Arie}}]{SLL14}
\bibinfo{author}{\bibfnamefont{R.}~\bibnamefont{Shiloh}},
  \bibinfo{author}{\bibfnamefont{Y.}~\bibnamefont{Lereah}},
  \bibinfo{author}{\bibfnamefont{Y.}~\bibnamefont{Lilach}}, \bibnamefont{and}
  \bibinfo{author}{\bibfnamefont{A.}~\bibnamefont{Arie}},
  \bibinfo{journal}{Ultramicroscopy} \textbf{\bibinfo{volume}{144}},
  \bibinfo{pages}{26} (\bibinfo{year}{2014}).

\bibitem[{\citenamefont{Schattschneider
  et~al.}(2012)\citenamefont{Schattschneider, {St\"{o}ger-Pollach}, and
  Verbeeck}}]{SSV12}
\bibinfo{author}{\bibfnamefont{P.}~\bibnamefont{Schattschneider}},
  \bibinfo{author}{\bibfnamefont{M.}~\bibnamefont{{St\"{o}ger-Pollach}}},
  \bibnamefont{and} \bibinfo{author}{\bibfnamefont{J.}~\bibnamefont{Verbeeck}},
  \bibinfo{journal}{Phys.\ Rev.\ Lett.} \textbf{\bibinfo{volume}{109}},
  \bibinfo{pages}{084801} (\bibinfo{year}{2012}).

\bibitem[{\citenamefont{Clark et~al.}(2013)\citenamefont{Clark, B\'ech\'e,
  Guzzinati, Lubk, Mazilu, Van~Boxem, and Verbeeck}}]{CBK13}
\bibinfo{author}{\bibfnamefont{L.}~\bibnamefont{Clark}},
  \bibinfo{author}{\bibfnamefont{A.}~\bibnamefont{B\'ech\'e}},
  \bibinfo{author}{\bibfnamefont{G.}~\bibnamefont{Guzzinati}},
  \bibinfo{author}{\bibfnamefont{A.}~\bibnamefont{Lubk}},
  \bibinfo{author}{\bibfnamefont{M.}~\bibnamefont{Mazilu}},
  \bibinfo{author}{\bibfnamefont{R.}~\bibnamefont{Van~Boxem}},
  \bibnamefont{and} \bibinfo{author}{\bibfnamefont{J.}~\bibnamefont{Verbeeck}},
  \bibinfo{journal}{Phys.\ Rev.\ Lett.} \textbf{\bibinfo{volume}{111}},
  \bibinfo{pages}{064801} (\bibinfo{year}{2013}).

\bibitem[{\citenamefont{Verbeeck et~al.}(2018)\citenamefont{Verbeeck,
  B{\'e}ch{\'e}, M{\"u}ller-Caspary, Guzzinati, Luong, and Den~Hertog}}]{VBM18}
\bibinfo{author}{\bibfnamefont{J.}~\bibnamefont{Verbeeck}},
  \bibinfo{author}{\bibfnamefont{A.}~\bibnamefont{B{\'e}ch{\'e}}},
  \bibinfo{author}{\bibfnamefont{K.}~\bibnamefont{M{\"u}ller-Caspary}},
  \bibinfo{author}{\bibfnamefont{G.}~\bibnamefont{Guzzinati}},
  \bibinfo{author}{\bibfnamefont{M.~A.} \bibnamefont{Luong}}, \bibnamefont{and}
  \bibinfo{author}{\bibfnamefont{M.}~\bibnamefont{Den~Hertog}},
  \bibinfo{journal}{Ultramicroscopy} \textbf{\bibinfo{volume}{190}},
  \bibinfo{pages}{58} (\bibinfo{year}{2018}).

\bibitem[{\citenamefont{Barwick et~al.}(2009)\citenamefont{Barwick, Flannigan,
  and Zewail}}]{BFZ09}
\bibinfo{author}{\bibfnamefont{B.}~\bibnamefont{Barwick}},
  \bibinfo{author}{\bibfnamefont{D.~J.} \bibnamefont{Flannigan}},
  \bibnamefont{and} \bibinfo{author}{\bibfnamefont{A.~H.}
  \bibnamefont{Zewail}}, \bibinfo{journal}{Nature}
  \textbf{\bibinfo{volume}{462}}, \bibinfo{pages}{902} (\bibinfo{year}{2009}).

\bibitem[{\citenamefont{Kirchner et~al.}(2014)\citenamefont{Kirchner, Gliserin,
  Krausz, and Baum}}]{KGK14}
\bibinfo{author}{\bibfnamefont{F.~O.} \bibnamefont{Kirchner}},
  \bibinfo{author}{\bibfnamefont{A.}~\bibnamefont{Gliserin}},
  \bibinfo{author}{\bibfnamefont{F.}~\bibnamefont{Krausz}}, \bibnamefont{and}
  \bibinfo{author}{\bibfnamefont{P.}~\bibnamefont{Baum}},
  \bibinfo{journal}{Nat.\ Photon.} \textbf{\bibinfo{volume}{8}},
  \bibinfo{pages}{52} (\bibinfo{year}{2014}).

\bibitem[{\citenamefont{Piazza et~al.}(2015)\citenamefont{Piazza, Lummen,
  {Qui\~{n}onez}, Murooka, Reed, Barwick, and Carbone}}]{PLQ15}
\bibinfo{author}{\bibfnamefont{L.}~\bibnamefont{Piazza}},
  \bibinfo{author}{\bibfnamefont{T.~T.~A.} \bibnamefont{Lummen}},
  \bibinfo{author}{\bibfnamefont{E.}~\bibnamefont{{Qui\~{n}onez}}},
  \bibinfo{author}{\bibfnamefont{Y.}~\bibnamefont{Murooka}},
  \bibinfo{author}{\bibfnamefont{B.}~\bibnamefont{Reed}},
  \bibinfo{author}{\bibfnamefont{B.}~\bibnamefont{Barwick}}, \bibnamefont{and}
  \bibinfo{author}{\bibfnamefont{F.}~\bibnamefont{Carbone}},
  \bibinfo{journal}{Nat.\ Commun.} \textbf{\bibinfo{volume}{6}},
  \bibinfo{pages}{6407} (\bibinfo{year}{2015}).

\bibitem[{\citenamefont{Feist et~al.}(2015)\citenamefont{Feist, Echternkamp,
  Schauss, Yalunin, Sch\"afer, and Ropers}}]{FES15}
\bibinfo{author}{\bibfnamefont{A.}~\bibnamefont{Feist}},
  \bibinfo{author}{\bibfnamefont{K.~E.} \bibnamefont{Echternkamp}},
  \bibinfo{author}{\bibfnamefont{J.}~\bibnamefont{Schauss}},
  \bibinfo{author}{\bibfnamefont{S.~V.} \bibnamefont{Yalunin}},
  \bibinfo{author}{\bibfnamefont{S.}~\bibnamefont{Sch\"afer}},
  \bibnamefont{and} \bibinfo{author}{\bibfnamefont{C.}~\bibnamefont{Ropers}},
  \bibinfo{journal}{Nature} \textbf{\bibinfo{volume}{521}},
  \bibinfo{pages}{200} (\bibinfo{year}{2015}).

\bibitem[{\citenamefont{Lummen et~al.}(2016)\citenamefont{Lummen, Lamb,
  Berruto, LaGrange, Negro, {Garc\'{\i}a de Abajo}, McGrouther, Barwick, and
  Carbone}}]{paper282}
\bibinfo{author}{\bibfnamefont{T.~T.~A.} \bibnamefont{Lummen}},
  \bibinfo{author}{\bibfnamefont{R.~J.} \bibnamefont{Lamb}},
  \bibinfo{author}{\bibfnamefont{G.}~\bibnamefont{Berruto}},
  \bibinfo{author}{\bibfnamefont{T.}~\bibnamefont{LaGrange}},
  \bibinfo{author}{\bibfnamefont{L.~D.} \bibnamefont{Negro}},
  \bibinfo{author}{\bibfnamefont{F.~J.} \bibnamefont{{Garc\'{\i}a de Abajo}}},
  \bibinfo{author}{\bibfnamefont{D.}~\bibnamefont{McGrouther}},
  \bibinfo{author}{\bibfnamefont{B.}~\bibnamefont{Barwick}}, \bibnamefont{and}
  \bibinfo{author}{\bibfnamefont{F.}~\bibnamefont{Carbone}},
  \bibinfo{journal}{Nat.\ Commun.} \textbf{\bibinfo{volume}{7}},
  \bibinfo{pages}{13156} (\bibinfo{year}{2016}).

\bibitem[{\citenamefont{Echternkamp et~al.}(2016)\citenamefont{Echternkamp,
  Feist, Sch\"{a}fer, and Ropers}}]{EFS16}
\bibinfo{author}{\bibfnamefont{K.~E.} \bibnamefont{Echternkamp}},
  \bibinfo{author}{\bibfnamefont{A.}~\bibnamefont{Feist}},
  \bibinfo{author}{\bibfnamefont{S.}~\bibnamefont{Sch\"{a}fer}},
  \bibnamefont{and} \bibinfo{author}{\bibfnamefont{C.}~\bibnamefont{Ropers}},
  \bibinfo{journal}{Nat.\ Phys.} \textbf{\bibinfo{volume}{12}},
  \bibinfo{pages}{1000} (\bibinfo{year}{2016}).

\bibitem[{\citenamefont{Kealhofer et~al.}(2016)\citenamefont{Kealhofer,
  Schneider, Ehberger, Ryabov, Krausz, and Baum}}]{KSE16}
\bibinfo{author}{\bibfnamefont{C.}~\bibnamefont{Kealhofer}},
  \bibinfo{author}{\bibfnamefont{W.}~\bibnamefont{Schneider}},
  \bibinfo{author}{\bibfnamefont{D.}~\bibnamefont{Ehberger}},
  \bibinfo{author}{\bibfnamefont{A.}~\bibnamefont{Ryabov}},
  \bibinfo{author}{\bibfnamefont{F.}~\bibnamefont{Krausz}}, \bibnamefont{and}
  \bibinfo{author}{\bibfnamefont{P.}~\bibnamefont{Baum}},
  \bibinfo{journal}{Science} \textbf{\bibinfo{volume}{352}},
  \bibinfo{pages}{429} (\bibinfo{year}{2016}).

\bibitem[{\citenamefont{Ryabov and Baum}(2016)}]{RB16}
\bibinfo{author}{\bibfnamefont{A.}~\bibnamefont{Ryabov}} \bibnamefont{and}
  \bibinfo{author}{\bibfnamefont{P.}~\bibnamefont{Baum}},
  \bibinfo{journal}{Science} \textbf{\bibinfo{volume}{353}},
  \bibinfo{pages}{374} (\bibinfo{year}{2016}).

\bibitem[{\citenamefont{Vanacore et~al.}(2016)\citenamefont{Vanacore,
  Fitzpatrick, and Zewail}}]{VFZ16}
\bibinfo{author}{\bibfnamefont{G.~M.} \bibnamefont{Vanacore}},
  \bibinfo{author}{\bibfnamefont{A.~W.~P.} \bibnamefont{Fitzpatrick}},
  \bibnamefont{and} \bibinfo{author}{\bibfnamefont{A.~H.}
  \bibnamefont{Zewail}}, \bibinfo{journal}{Nano\ Today}
  \textbf{\bibinfo{volume}{11}}, \bibinfo{pages}{228} (\bibinfo{year}{2016}).

\bibitem[{\citenamefont{Morimoto and Baum}(2017)}]{MB17}
\bibinfo{author}{\bibfnamefont{Y.}~\bibnamefont{Morimoto}} \bibnamefont{and}
  \bibinfo{author}{\bibfnamefont{P.}~\bibnamefont{Baum}},
  \bibinfo{journal}{Nat. Phys.} \textbf{\bibinfo{volume}{14}},
  \bibinfo{pages}{252} (\bibinfo{year}{2017}).

\bibitem[{\citenamefont{Koz\'ak et~al.}(2017)\citenamefont{Koz\'ak, McNeur,
  Leedle, Deng, Sch\"onenberger, Ruehl, Hartl, Harris, Byer, and
  Hommelhoff}}]{KML17}
\bibinfo{author}{\bibfnamefont{M.}~\bibnamefont{Koz\'ak}},
  \bibinfo{author}{\bibfnamefont{J.}~\bibnamefont{McNeur}},
  \bibinfo{author}{\bibfnamefont{K.~J.} \bibnamefont{Leedle}},
  \bibinfo{author}{\bibfnamefont{H.}~\bibnamefont{Deng}},
  \bibinfo{author}{\bibfnamefont{N.}~\bibnamefont{Sch\"onenberger}},
  \bibinfo{author}{\bibfnamefont{A.}~\bibnamefont{Ruehl}},
  \bibinfo{author}{\bibfnamefont{I.}~\bibnamefont{Hartl}},
  \bibinfo{author}{\bibfnamefont{J.~S.} \bibnamefont{Harris}},
  \bibinfo{author}{\bibfnamefont{R.~L.} \bibnamefont{Byer}}, \bibnamefont{and}
  \bibinfo{author}{\bibfnamefont{P.}~\bibnamefont{Hommelhoff}},
  \bibinfo{journal}{Nat.\ Commun.} \textbf{\bibinfo{volume}{8}},
  \bibinfo{pages}{14342} (\bibinfo{year}{2017}).

\bibitem[{\citenamefont{Feist et~al.}(2017)\citenamefont{Feist, Bach,
  N.~Rubiano~{da Silva}, M\"{a}ller, Priebe, Domr\"{a}se, Gatzmann, Rost,
  Schauss, Strauch et~al.}}]{FBR17}
\bibinfo{author}{\bibfnamefont{A.}~\bibnamefont{Feist}},
  \bibinfo{author}{\bibfnamefont{N.}~\bibnamefont{Bach}},
  \bibinfo{author}{\bibfnamefont{T.~D.} \bibnamefont{N.~Rubiano~{da Silva}}},
  \bibinfo{author}{\bibfnamefont{M.}~\bibnamefont{M\"{a}ller}},
  \bibinfo{author}{\bibfnamefont{K.~E.} \bibnamefont{Priebe}},
  \bibinfo{author}{\bibfnamefont{T.}~\bibnamefont{Domr\"{a}se}},
  \bibinfo{author}{\bibfnamefont{J.~G.} \bibnamefont{Gatzmann}},
  \bibinfo{author}{\bibfnamefont{S.}~\bibnamefont{Rost}},
  \bibinfo{author}{\bibfnamefont{J.}~\bibnamefont{Schauss}},
  \bibinfo{author}{\bibfnamefont{S.}~\bibnamefont{Strauch}},
  \bibnamefont{et~al.}, \bibinfo{journal}{Ultramicroscopy}
  \textbf{\bibinfo{volume}{176}}, \bibinfo{pages}{63} (\bibinfo{year}{2017}).

\bibitem[{\citenamefont{Priebe et~al.}(2017)\citenamefont{Priebe, Rathje,
  Yalunin, Hohage, Feist, Sch\"{a}fer, and Ropers}}]{PRY17}
\bibinfo{author}{\bibfnamefont{K.~E.} \bibnamefont{Priebe}},
  \bibinfo{author}{\bibfnamefont{C.}~\bibnamefont{Rathje}},
  \bibinfo{author}{\bibfnamefont{S.~V.} \bibnamefont{Yalunin}},
  \bibinfo{author}{\bibfnamefont{T.}~\bibnamefont{Hohage}},
  \bibinfo{author}{\bibfnamefont{A.}~\bibnamefont{Feist}},
  \bibinfo{author}{\bibfnamefont{S.}~\bibnamefont{Sch\"{a}fer}},
  \bibnamefont{and} \bibinfo{author}{\bibfnamefont{C.}~\bibnamefont{Ropers}},
  \bibinfo{journal}{Nat.\ Photon.} \textbf{\bibinfo{volume}{11}},
  \bibinfo{pages}{793} (\bibinfo{year}{2017}).

\bibitem[{\citenamefont{Pomarico et~al.}(2018)\citenamefont{Pomarico, Madan,
  Berruto, Vanacore, Wang, Kaminer, {Garc\'{\i}a de Abajo}, and
  Carbone}}]{paper306}
\bibinfo{author}{\bibfnamefont{E.}~\bibnamefont{Pomarico}},
  \bibinfo{author}{\bibfnamefont{I.}~\bibnamefont{Madan}},
  \bibinfo{author}{\bibfnamefont{G.}~\bibnamefont{Berruto}},
  \bibinfo{author}{\bibfnamefont{G.~M.} \bibnamefont{Vanacore}},
  \bibinfo{author}{\bibfnamefont{K.}~\bibnamefont{Wang}},
  \bibinfo{author}{\bibfnamefont{I.}~\bibnamefont{Kaminer}},
  \bibinfo{author}{\bibfnamefont{F.~J.} \bibnamefont{{Garc\'{\i}a de Abajo}}},
  \bibnamefont{and} \bibinfo{author}{\bibfnamefont{F.}~\bibnamefont{Carbone}},
  \bibinfo{journal}{ACS\ Photonics} \textbf{\bibinfo{volume}{5}},
  \bibinfo{pages}{759} (\bibinfo{year}{2018}).

\bibitem[{\citenamefont{Vanacore et~al.}(2018)\citenamefont{Vanacore, Madan,
  Berruto, Wang, Pomarico, Lamb, McGrouther, Kaminer, Barwick, {Garc\'{\i}a de
  Abajo} et~al.}}]{paper311}
\bibinfo{author}{\bibfnamefont{G.~M.} \bibnamefont{Vanacore}},
  \bibinfo{author}{\bibfnamefont{I.}~\bibnamefont{Madan}},
  \bibinfo{author}{\bibfnamefont{G.}~\bibnamefont{Berruto}},
  \bibinfo{author}{\bibfnamefont{K.}~\bibnamefont{Wang}},
  \bibinfo{author}{\bibfnamefont{E.}~\bibnamefont{Pomarico}},
  \bibinfo{author}{\bibfnamefont{R.~J.} \bibnamefont{Lamb}},
  \bibinfo{author}{\bibfnamefont{D.}~\bibnamefont{McGrouther}},
  \bibinfo{author}{\bibfnamefont{I.}~\bibnamefont{Kaminer}},
  \bibinfo{author}{\bibfnamefont{B.}~\bibnamefont{Barwick}},
  \bibinfo{author}{\bibfnamefont{F.~J.} \bibnamefont{{Garc\'{\i}a de Abajo}}},
  \bibnamefont{et~al.}, \bibinfo{journal}{Nat.\ Commun.}
  \textbf{\bibinfo{volume}{9}}, \bibinfo{pages}{2694} (\bibinfo{year}{2018}).

\bibitem[{\citenamefont{Morimoto and Baum}(2018)}]{MB18}
\bibinfo{author}{\bibfnamefont{Y.}~\bibnamefont{Morimoto}} \bibnamefont{and}
  \bibinfo{author}{\bibfnamefont{P.}~\bibnamefont{Baum}},
  \bibinfo{journal}{Phys.\ Rev.\ A} \textbf{\bibinfo{volume}{97}},
  \bibinfo{pages}{033815} (\bibinfo{year}{2018}).

\bibitem[{\citenamefont{Vanacore et~al.}(2019)\citenamefont{Vanacore, Berruto,
  Madan, Pomarico, Biagioni, Lamb, McGrouther, Reinhardt, Kaminer, Barwick
  et~al.}}]{paper332}
\bibinfo{author}{\bibfnamefont{G.~M.} \bibnamefont{Vanacore}},
  \bibinfo{author}{\bibfnamefont{G.}~\bibnamefont{Berruto}},
  \bibinfo{author}{\bibfnamefont{I.}~\bibnamefont{Madan}},
  \bibinfo{author}{\bibfnamefont{E.}~\bibnamefont{Pomarico}},
  \bibinfo{author}{\bibfnamefont{P.}~\bibnamefont{Biagioni}},
  \bibinfo{author}{\bibfnamefont{R.~J.} \bibnamefont{Lamb}},
  \bibinfo{author}{\bibfnamefont{D.}~\bibnamefont{McGrouther}},
  \bibinfo{author}{\bibfnamefont{O.}~\bibnamefont{Reinhardt}},
  \bibinfo{author}{\bibfnamefont{I.}~\bibnamefont{Kaminer}},
  \bibinfo{author}{\bibfnamefont{B.}~\bibnamefont{Barwick}},
  \bibnamefont{et~al.}, \bibinfo{journal}{Nat.\ Mater.}
  \textbf{\bibinfo{volume}{18}}, \bibinfo{pages}{573} (\bibinfo{year}{2019}).

\bibitem[{\citenamefont{Wang et~al.}(2019)\citenamefont{Wang, Dahan, Shentcis,
  Kauffmann, Tsesses, , and Kaminer}}]{WDS19}
\bibinfo{author}{\bibfnamefont{K.}~\bibnamefont{Wang}},
  \bibinfo{author}{\bibfnamefont{R.}~\bibnamefont{Dahan}},
  \bibinfo{author}{\bibfnamefont{M.}~\bibnamefont{Shentcis}},
  \bibinfo{author}{\bibfnamefont{Y.}~\bibnamefont{Kauffmann}},
  \bibinfo{author}{\bibfnamefont{S.}~\bibnamefont{Tsesses}}, ,
  \bibnamefont{and} \bibinfo{author}{\bibfnamefont{I.}~\bibnamefont{Kaminer}},
  p. \bibinfo{pages}{arXiv:1908.06206} (\bibinfo{year}{2019}).

\bibitem[{\citenamefont{Dahan et~al.}(2019)\citenamefont{Dahan, Nehemia,
  Shentcis, Reinhardt, Adiv, Wang, Beer, Kurman, Shi, Lynch et~al.}}]{DNS19}
\bibinfo{author}{\bibfnamefont{R.}~\bibnamefont{Dahan}},
  \bibinfo{author}{\bibfnamefont{S.}~\bibnamefont{Nehemia}},
  \bibinfo{author}{\bibfnamefont{M.}~\bibnamefont{Shentcis}},
  \bibinfo{author}{\bibfnamefont{O.}~\bibnamefont{Reinhardt}},
  \bibinfo{author}{\bibfnamefont{Y.}~\bibnamefont{Adiv}},
  \bibinfo{author}{\bibfnamefont{K.}~\bibnamefont{Wang}},
  \bibinfo{author}{\bibfnamefont{O.}~\bibnamefont{Beer}},
  \bibinfo{author}{\bibfnamefont{Y.}~\bibnamefont{Kurman}},
  \bibinfo{author}{\bibfnamefont{X.}~\bibnamefont{Shi}},
  \bibinfo{author}{\bibfnamefont{M.~H.} \bibnamefont{Lynch}},
  \bibnamefont{et~al.}, p. \bibinfo{pages}{arXiv:1909.00757}
  (\bibinfo{year}{2019}).

\bibitem[{\citenamefont{Kfir et~al.}(2019)\citenamefont{Kfir,
  Louren\c{c}o-Martins, Storeck, Sivis, Harvey, Kippenberg, Feist, and
  Ropers}}]{KLS19}
\bibinfo{author}{\bibfnamefont{O.}~\bibnamefont{Kfir}},
  \bibinfo{author}{\bibfnamefont{H.}~\bibnamefont{Louren\c{c}o-Martins}},
  \bibinfo{author}{\bibfnamefont{G.}~\bibnamefont{Storeck}},
  \bibinfo{author}{\bibfnamefont{M.}~\bibnamefont{Sivis}},
  \bibinfo{author}{\bibfnamefont{T.~R.} \bibnamefont{Harvey}},
  \bibinfo{author}{\bibfnamefont{T.~J.} \bibnamefont{Kippenberg}},
  \bibinfo{author}{\bibfnamefont{A.}~\bibnamefont{Feist}}, \bibnamefont{and}
  \bibinfo{author}{\bibfnamefont{C.}~\bibnamefont{Ropers}}, p.
  \bibinfo{pages}{arXiv:1910.09540} (\bibinfo{year}{2019}).

\bibitem[{\citenamefont{{Garc\'{\i}a de Abajo}
  et~al.}(2010)\citenamefont{{Garc\'{\i}a de Abajo}, {Asenjo Garcia}, and
  Kociak}}]{paper151}
\bibinfo{author}{\bibfnamefont{F.~J.} \bibnamefont{{Garc\'{\i}a de Abajo}}},
  \bibinfo{author}{\bibfnamefont{A.}~\bibnamefont{{Asenjo Garcia}}},
  \bibnamefont{and} \bibinfo{author}{\bibfnamefont{M.}~\bibnamefont{Kociak}},
  \bibinfo{journal}{Nano\ Lett.} \textbf{\bibinfo{volume}{10}},
  \bibinfo{pages}{1859} (\bibinfo{year}{2010}).

\bibitem[{\citenamefont{Park et~al.}(2010)\citenamefont{Park, Lin, and
  Zewail}}]{PLZ10}
\bibinfo{author}{\bibfnamefont{S.~T.} \bibnamefont{Park}},
  \bibinfo{author}{\bibfnamefont{M.}~\bibnamefont{Lin}}, \bibnamefont{and}
  \bibinfo{author}{\bibfnamefont{A.~H.} \bibnamefont{Zewail}},
  \bibinfo{journal}{New\ J.\ Phys.} \textbf{\bibinfo{volume}{12}},
  \bibinfo{pages}{123028} (\bibinfo{year}{2010}).

\bibitem[{\citenamefont{Park and Zewail}(2012)}]{PZ12}
\bibinfo{author}{\bibfnamefont{S.~T.} \bibnamefont{Park}} \bibnamefont{and}
  \bibinfo{author}{\bibfnamefont{A.~H.} \bibnamefont{Zewail}},
  \bibinfo{journal}{J.\ Phys.\ Chem.\ A} \textbf{\bibinfo{volume}{116}},
  \bibinfo{pages}{11128} (\bibinfo{year}{2012}).

\bibitem[{\citenamefont{{Garc\'{\i}a de Abajo}
  et~al.}(2016)\citenamefont{{Garc\'{\i}a de Abajo}, Barwick, and
  Carbone}}]{paper272}
\bibinfo{author}{\bibfnamefont{F.~J.} \bibnamefont{{Garc\'{\i}a de Abajo}}},
  \bibinfo{author}{\bibfnamefont{B.}~\bibnamefont{Barwick}}, \bibnamefont{and}
  \bibinfo{author}{\bibfnamefont{F.}~\bibnamefont{Carbone}},
  \bibinfo{journal}{Phys.\ Rev.\ B} \textbf{\bibinfo{volume}{94}},
  \bibinfo{pages}{041404(R)} (\bibinfo{year}{2016}).

\bibitem[{\citenamefont{Cai et~al.}(2018)\citenamefont{Cai, Reinhardt, Kaminer,
  and {Garc\'{\i}a de Abajo}}}]{paper312}
\bibinfo{author}{\bibfnamefont{W.}~\bibnamefont{Cai}},
  \bibinfo{author}{\bibfnamefont{O.}~\bibnamefont{Reinhardt}},
  \bibinfo{author}{\bibfnamefont{I.}~\bibnamefont{Kaminer}}, \bibnamefont{and}
  \bibinfo{author}{\bibfnamefont{F.~J.} \bibnamefont{{Garc\'{\i}a de Abajo}}},
  \bibinfo{journal}{Phys.\ Rev.\ B} \textbf{\bibinfo{volume}{98}},
  \bibinfo{pages}{045424} (\bibinfo{year}{2018}).

\bibitem[{\citenamefont{{Di Giulio} et~al.}(2019)\citenamefont{{Di Giulio},
  Kociak, and {Garc\'{\i}a de Abajo}}}]{paper339}
\bibinfo{author}{\bibfnamefont{V.}~\bibnamefont{{Di Giulio}}},
  \bibinfo{author}{\bibfnamefont{M.}~\bibnamefont{Kociak}}, \bibnamefont{and}
  \bibinfo{author}{\bibfnamefont{F.~J.} \bibnamefont{{Garc\'{\i}a de Abajo}}},
  \bibinfo{journal}{Optica} \textbf{\bibinfo{volume}{6}}, \bibinfo{pages}{1524}
  (\bibinfo{year}{2019}).

\bibitem[{\citenamefont{Kfir}(2019)}]{K19}
\bibinfo{author}{\bibfnamefont{O.}~\bibnamefont{Kfir}},
  \bibinfo{journal}{Phys.\ Rev.\ Lett.} \textbf{\bibinfo{volume}{123}},
  \bibinfo{pages}{103602} (\bibinfo{year}{2019}).

\bibitem[{\citenamefont{Reinhardt et~al.}(2019)\citenamefont{Reinhardt, Mechel,
  Lynch, and Kaminer}}]{RML19}
\bibinfo{author}{\bibfnamefont{O.}~\bibnamefont{Reinhardt}},
  \bibinfo{author}{\bibfnamefont{C.}~\bibnamefont{Mechel}},
  \bibinfo{author}{\bibfnamefont{M.}~\bibnamefont{Lynch}}, \bibnamefont{and}
  \bibinfo{author}{\bibfnamefont{I.}~\bibnamefont{Kaminer}},
  \emph{\bibinfo{title}{Free-electron qubits}} (\bibinfo{year}{2019}),
  \eprint{1907.10281}.

\bibitem[{\citenamefont{Allen et~al.}(2001)\citenamefont{Allen, Oxley, and
  Paganin}}]{AOP01}
\bibinfo{author}{\bibfnamefont{L.~J.} \bibnamefont{Allen}},
  \bibinfo{author}{\bibfnamefont{M.~P.} \bibnamefont{Oxley}}, \bibnamefont{and}
  \bibinfo{author}{\bibfnamefont{D.}~\bibnamefont{Paganin}},
  \bibinfo{journal}{Phys.\ Rev.\ Lett.} \textbf{\bibinfo{volume}{87}},
  \bibinfo{pages}{123902} (\bibinfo{year}{2001}).

\bibitem[{\citenamefont{Hawkes et~al.}(1995)\citenamefont{Hawkes, Lencov{\'a},
  and Lenc}}]{HLL1995}
\bibinfo{author}{\bibfnamefont{P.}~\bibnamefont{Hawkes}},
  \bibinfo{author}{\bibfnamefont{B.}~\bibnamefont{Lencov{\'a}}},
  \bibnamefont{and} \bibinfo{author}{\bibfnamefont{M.}~\bibnamefont{Lenc}},
  \bibinfo{journal}{J.\ Microsc.} \textbf{\bibinfo{volume}{179}},
  \bibinfo{pages}{145} (\bibinfo{year}{1995}).

\bibitem[{\citenamefont{Egerton}(2005)}]{E05}
\bibinfo{author}{\bibfnamefont{R.~F.} \bibnamefont{Egerton}},
  \emph{\bibinfo{title}{Physical Principles of Electron Microscopy: An
  Introduction to {TEM}, {SEM}, and {AEM}}} (\bibinfo{publisher}{Springer},
  \bibinfo{year}{2005}).

\bibitem[{\citenamefont{Paganin et~al.}(2018)\citenamefont{Paganin, Petersen,
  and Beltran}}]{PPB18}
\bibinfo{author}{\bibfnamefont{D.~M.} \bibnamefont{Paganin}},
  \bibinfo{author}{\bibfnamefont{T.~C.} \bibnamefont{Petersen}},
  \bibnamefont{and} \bibinfo{author}{\bibfnamefont{M.~A.}
  \bibnamefont{Beltran}}, \bibinfo{journal}{Phys.\ Rev.\ A}
  \textbf{\bibinfo{volume}{97}}, \bibinfo{pages}{023835}
  (\bibinfo{year}{2018}).

\bibitem[{\citenamefont{Abramowitz and Stegun}(1972)}]{AS1972}
\bibinfo{author}{\bibfnamefont{M.}~\bibnamefont{Abramowitz}} \bibnamefont{and}
  \bibinfo{author}{\bibfnamefont{I.~A.} \bibnamefont{Stegun}},
  \emph{\bibinfo{title}{Handbook of Mathematical Functions}}
  (\bibinfo{publisher}{Dover}, \bibinfo{address}{New York},
  \bibinfo{year}{1972}).

\bibitem[{\citenamefont{Watson}(1944)}]{W1944}
\bibinfo{author}{\bibfnamefont{G.~N.} \bibnamefont{Watson}},
  \emph{\bibinfo{title}{A Treatise on the Theory of Bessel Functions}}
  (\bibinfo{publisher}{Cambridge}, \bibinfo{address}{Cambridge University
  Press}, \bibinfo{year}{1944}).

\end{thebibliography}

\end{document}